\documentclass[runningheads]{llncs}
\usepackage[T1]{fontenc}
\usepackage{graphicx}
\usepackage{amssymb}
\usepackage{amsmath}
\usepackage[utf8]{inputenc}
\usepackage{mathtools}
\usepackage{stmaryrd}
\usepackage{hyperref}
\usepackage{color}

\usepackage[a4paper]{geometry}

\begin{document}
\title{Abstraction Logic}
\subtitle{A New Foundation for (Computer) Mathematics}
%
%
\author{Steven Obua}

%
%
\institute{\email{obua@practal.com}}
\maketitle              
\begin{abstract}
Abstraction logic is a new logic, serving as a foundation of mathematics. It combines features of both predicate logic and higher-order logic: 
abstraction logic can be viewed both as higher-order logic minus static types as well as predicate logic plus operators and variable binding. We argue
that abstraction logic is the best foundational logic possible because it maximises both simplicity and practical expressivity. 
This argument is supported by the observation that abstraction logic has simpler terms and a simpler notion of proof than all other general logics. At the same time,
abstraction logic can formalise both intuitionistic and classical abstraction logic, and is sound and complete for these logics and all other logics
extending deduction logic with equality. 
\keywords{Interactive Theorem Proving \and Algebraic Semantics \and Predicate Logic \and Higher-Order Logic \and Operators 
\and Variable Binding \and Undefinedness \and Practical Types \and Practal}
\end{abstract}
\newcommand{\true}{\top}
\newcommand{\false}{\bot}
\newcommand{\fail}{\Finv}
\newcommand{\imp}{\Rightarrow}
\newcommand{\all}{\forall}
\newcommand{\ex}{\exists}
\newcommand{\exu}{\exists_1}
\newcommand{\techterm}[1]{\textbf{#1}}
\newcommand{\deftechterm}[1]{\textbf{\textit{#1}}}
\newcommand{\univ}{\mathcal{U}}
\newcommand{\size}[1]{{|#1|}}
\newcommand{\sig}{{\mathfrak{S}}}
\newcommand{\logic}{{\mathcal{L}}}
\newcommand{\axioms}{{\mathbb{A}}}
\newcommand{\val}[1]{\emph{#1}}
\newcommand{\ctrue}{\true}
\newcommand{\cfalse}{\false}
\newcommand{\cimplies}{\imp}
\newcommand{\cforall}{\all}
\newcommand{\cexists}{\exists}
\newcommand{\eq}{=}
\newcommand{\ceq}{\eq}
\newcommand{\cnot}{\neg}
\newcommand{\cor}{\vee}
\newcommand{\cand}{\wedge}
\newcommand{\cequiv}{\Leftrightarrow}
\newcommand{\absatz}{\vspace{0.1cm}\noindent}
\newcommand{\valueof}[1]{\llbracket #1 \rrbracket}
\newcommand{\appsubst}[2]{#2 / #1}
\newcommand{\valuationspace}{\mathcal{V}}
\newcommand{\algebra}{\mathfrak{A}}
\newcommand{\valid}{\models}
\newcommand{\derives}{\vdash}
\newcommand{\pAX}{\operatorname{AX}}
\newcommand{\pSUBST}{\operatorname{SUBST}}
\newcommand{\pALL}{\operatorname{ALL}}
\newcommand{\pMP}{\operatorname{MP}}
\newcommand{\timplies}{\text{$\imp$}}
\newcommand{\tall}{\text{$\all$}}
\newcommand{\ralg}{\mathfrak{R}}
\newcommand{\rclass}[1]{{[#1]_{\ralg}}}
\newcommand{\lax}[2]{$\text{#1}_{#2}$}

\newcommand{\opnat}{\operatorname{nat}}
\newcommand{\opsuc}{\operatorname{suc}}
\newcommand{\opzero}{\operatorname{zero}}
\newcommand{\opadd}{\operatorname{add}}
\newcommand{\opmul}{\operatorname{mul}}
\newcommand{\opnatapp}[1]{{({\opnat}.\, #1)}}
\newcommand{\opsucapp}[1]{{({\opsuc}.\, #1)}}
\newcommand{\opaddapp}[2]{{({\opadd}.\, #1\, #2)}}
\newcommand{\opmulapp}[2]{{({\opmul}.\, #1\, #2)}}

\newcommand{\emphabs}[1]{``$\operatorname{#1}$''}

\section{Introduction}
\techterm{Propositional logic} and \techterm{predicate logic} / \techterm{first-order logic} (FOL) are widely accepted as logical foundations of mathematics. 
Nevertheless, for the task of formulating and proving mathematical theorems on a computer, they have been found lacking for practical reasons, and are therefore usually replaced by some form of \techterm{higher-order logic}.
The main reason for the prevalence of higher-order logic in \techterm{interactive theorem proving} is not philosophical, but mundane: Predicate logic admits only a limited form of 
\deftechterm{variable binding}, namely as part of universal and existential quantification, 
while higher-order logic allows arbitrary variable binding via some variant of the \techterm{$\lambda$-calculus}. This means that it is not possible in FOL to directly represent
$$
\int_D F(x)\,dx
$$
as a term, because in $F(x)$ the variable $x$ is bound by the integral operator; but it is straightforward to do so in higher-order logic. 



\techterm{Abstraction logic} (AL) is a new logic I introduced last year in two articles~\cite{pal} and~\cite{al} I self-published on the web. It combines features of both predicate logic and simply-typed higher-order logic~(HOL)~\cite{church-hol}. Like HOL, it improves on predicate logic by properly supporting variable binding. Unlike HOL, it does not insist on dividing the \techterm{mathematical universe} into disjoint pieces labelled by static types. Just as in FOL, there is a single mathematical universe in which all mathematical objects live. Unlike for FOL, but like in HOL, \val{true} is a legitimate and even necessary element of the mathematical universe. Like FOL and HOL (with Henkin semantics), AL is sound and complete, assuming at least the axioms
of \techterm{deduction logic with equality}.

There is no consensus on whether there is a single best logic to use as a foundation of mathematics. There are many different logics, and they differ in how (much) mathematics can be expressed in them. On one end of the spectrum one can find the opinion that there cannot be a single best logic, and that the best we can do is to develop methods~\cite{rabe-future} which exploit commonalities among the plethora of logics. On the other end of the spectrum there is the opinion, championed by Quine~\cite{quine}, that FOL is the proper foundation of mathematics. In Quine's opinion, there is a clear boundary between logic and mathematics, and other logics than FOL are not really logic, but simply \emph{mathematics dressed up as logic}. He considers \techterm{second-order logic} just as another way of formulating \techterm{set theory}, which in his opinion does not qualify as logic, but is mathematics. 

Similarly, one can view HOL as the result of turning the mathematical theory of \techterm{simple types} into a logic. Adopting HOL as a foundation comes therefore with the consequence of also adopting this particular mathematical theory. If variable binding is all you wanted, you now bought much more than you bargained for!

Abstraction logic on the other hand is \emph{pure logic}. I believe that Quine has got it almost right. But it is not FOL which is the one true logic. It cannot be, as it does not support variable binding. Instead, AL is. Of course, I would not go so far as to deny all other logics their status as logic. But I believe that all of them, or at least all of those that are commonly used as foundations, can be formulated in AL as \deftechterm{mathematical theories}, where a mathematical theory is just a collection of constants (called abstractions) and axioms. 
This expressive power of AL comes from the fact that its semantics is a generalisation of algebraic semantics: the models of AL are generalisations of the algebraic models used by Rasiowa for nonstandard propositional logics~\cite{rasiowa}. 

Furthermore, abstraction logic is \emph{simpler} than all other general logics I know of. Its terms are simpler, and its notion of proof is simpler, too. In fact, it seems impossible to conceive
of simpler terms, or of a simpler notion of proof, at least for a general logic.
Therefore, were it not for tradition, AL would be easier to understand than all other general logics. 
With AL, there is no need to philosophically analyse what propositions and predicates \emph{really} are.
There is simply the mathematical universe, together with the ability of combining the values and operations of that universe to form other values of the universe.

Observing the purity and simplicity of AL together with its practical expressiveness, I find it hard to escape the conclusion that AL is indeed the \emph{best} logic.  

\absatz\textbf{Overview.}
The paper largely follows~\cite{pal} in structure and content. We first set the stage for the objects that AL deals with. We quickly proceed to the syntax of AL, and introduce then a range of abstraction logics, including both \techterm{intuitionistic} and \techterm{classical} abstraction logic. After discussing formalisations of \techterm{Peano arithmetic} and 
\techterm{undefinedness} in AL, we briefly talk about the origins of AL. Next we define the \techterm{semantics} of AL, and its notion of \techterm{proof}. Subsequently \techterm{soundness}, \techterm{completeness} and \techterm{consistency} of AL are treated. After commenting on the relationship between abstraction logic and Rasiowa's algebraic approach to logic, I remark on further related work and conclude.

\section{The Mathematical Universe}
Abstraction logic presumes that all \deftechterm{values} we want to calculate with and reason about live in a \deftechterm{single mathematical universe $\univ$}. This is the same assumption that predicate logic makes. Unlike FOL, abstraction logic makes the additional assumption that there also is a value \val{true} in $\univ$. Apart from this, AL makes no a-priori demands on $\univ$. 

The assumption that there is a single universe is an important reason why AL is conceptually so simple. Any value can in principle be combined with any other value without any a-priori restrictions. This is also what makes \techterm{set theory} so attractive, but unlike set theory, abstraction logic has no ontological presuppositions other than that $\univ$ must contain \val{true}.

Values are combined via \emph{operations}. Values and operations are combined via \emph{operators}. 

An $n$-ary \deftechterm{operation}(of $\univ$) is a function taking $n$ arguments where each argument is a value in $\univ$, and returning a result that again is a value in $\univ$. 

An $n$-ary \deftechterm{operator} (of $\univ$) is a function taking $n$ arguments where each argument is either a value or an $m$-ary operation, depending on the argument position; $m$ is fixed for a given position. The result of applying an operator is still always a value. 

We consider any value to also be an $0$-ary operation, and any $n$-ary operation to also be an $n$-ary operator. 
If we want to emphasise that an operation is not a value we call that operation \deftechterm{proper}. Similarly, we call an operator \deftechterm{proper} if it is not an operation. 

Note that \emph{proper operations and operators are \textbf{not} part of the universe}. There are simply too many of them for this to be the case if $\univ$ contains more than one value, as we know due to Cantor's theorem. Therefore all attempts to include all operations and/or all operators in the universe are doomed.

A most important difference between abstraction logic and first-order logic is that while FOL only considers arbitrary operations (in the form of functions and relations / predicates), AL also allows the use of \emph{arbitrary operators}. The only proper operators allowed in FOL are $\forall$ and $\exists$.

\section{Abstraction and Shape}

An \deftechterm{abstraction} is a name, intended to denote an operator. A \deftechterm{signature} is a collection of abstractions, where each abstraction is associated with a \emph{shape}. A shape encodes both what kind of operator an abstraction can possibly denote, and how to interpret a \emph{term} formed by applying the abstraction. 

In general, a \deftechterm{shape} has the form
$(m; p_0, \ldots, p_{n-1})$
where $m$ is the \deftechterm{valence} and $n$ the \deftechterm{arity} of the shape. The $p_i$ are subsets of $\{0, \ldots, m-1\}$. To avoid \emph{degenerate} cases, we demand that the union of all $p_i$ must equal $\{0, \ldots, m-1\}$.

We say that the operator $o$ is \deftechterm{compatible} with the shape $(m; p_0, \ldots, p_{n-1})$ in all of these cases:
\begin{itemize}
\item The arity $n$ is zero (which implies $m = 0$, because we excluded degenerate shapes) and $o$ is a value.
\item The valence $m$ is zero, the arity $n$ is non-zero, and $o$ is an $n$-ary operation. Note that $o$ is then necessarily a \emph{proper} operation.
\item The valence $m$ is non-zero (which implies $n > 0$) and $o$ is an $n$-ary operator, such that 
  $o$ accepts in argument position $i$ a $\size{p_i}$-ary operation. 
  Here $\size{p_i}$ is the size of $p_i$. Note that $o$ is then necessarily a \emph{proper} operator.
\end{itemize}
In all other cases $o$ is not compatible with the shape.

Because degenerate shapes are excluded, a value is only compatible with shape $(0;)$, a unary operation must have shape $(0; \emptyset)$, a binary operation must have shape $(0; \emptyset, \emptyset)$, and a proper unary operator necessarily has shape $(1; \{0\})$.

An abstraction can only denote operators that are compatible with its shape.

\section{Syntax}

We assume an infinite pool $X$ of \deftechterm{variables}. 
A \deftechterm{term}, relative to a given signature, is either a \emph{variable application} or an \emph{abstraction application}. 

A \deftechterm{variable application} has the form
$$x[t_0, \ldots, t_{n-1}]$$
where $x$ is a variable, and the $t_i$ are themselves terms. Instead of $x[]$ we usually just write $x$. We say that $x$ \emph{occurs with \deftechterm{arity} $n$}. 

An \deftechterm{abstraction application} has the form
$$(a\, x_0 \ldots x_{m-1}.\, t_0 \ldots t_{n-1})$$
where $a$ is an abstraction belonging to the signature, $m$ is the \deftechterm{valence} of $a$, and $n$ is the \deftechterm{arity} of $a$. The $t_i$ are the terms to which the abstraction is being applied. The $x_j$ are distinct variables. 

Often, instead of $(a.)$ we just write $a$, instead of $(a.\, s\, t)$ we write $s\, a\, t$ when appropriate, and instead of $(a.\, t)$ we might write $a\, t$ or $t\, a$, depending on whether we consider $a$ to be a prefix or postfix unary operator. In these cases we might disambiguate via brackets $(\ldots)$ and the usual assumptions about associativity and precedence.

We consider the arity with which a variable occurs to be an implicit part of the variable name. This means that the two occurrences of the variable $x$ in $x[x]$ actually refer to two different variables. We could emphasise this by writing $x^1[x^0]$ instead, but we choose the simpler syntax, as no ambiguities can arise. Because of this convention, we do not need to distinguish between wellformed terms and non-wellformed terms. As long as abstraction applications use the correct arity and valence according to the signature, \emph{all terms are automatically wellformed}.

Consider the expression $\int_D x\,dx$. To represent it as an AL term $t$, we introduce an abstraction \emphabs{integral} of shape $(1; \emptyset, \{0\})$. Then $t$ is
$$
(\operatorname{integral} x.\, D\, x)
$$
On the other hand, in HOL the constant \emphabs{integral} has a type like 
$$\operatorname{integral} : (\mathbb{R} \rightarrow \operatorname{bool}) \rightarrow (\mathbb{R} \rightarrow \mathbb{R}) \rightarrow 
\mathbb{R}\operatorname{option}$$
You can see, we already had to make many upfront decisions just to declare the constant, and they probably were not the best ones. The HOL term would then look like this:
$$
\operatorname{integral}\, D\, (\lambda x.\, x)
$$
AL term and HOL term for $\int_D x\,dx$ seem not that different. But AL terms do not delegate variable bindings to a $\lambda$-construct, but instead make it a direct part of the constant. 
The difference is that
AL terms always denote an element of the mathematical universe, while HOL terms also include higher-order terms like $(\lambda x.\, x)$. This forces a typing discipline upon HOL to maintain consistency. There is no such pressure on AL.

\section{Logic}

A \deftechterm{logic signature} is a signature $\sig$ that contains at least three abstractions:
\begin{itemize}
\item The abstraction $\true$, intended to denote a value.
\item The abstraction $\imp$, intended to denote a binary operation.
\item The abstraction $\all$, intended to denote a proper unary operator. 
\end{itemize}

An \deftechterm{abstraction logic} is a logic signature $\sig$ together with a list $\axioms$ of terms called \deftechterm{axioms}. 

Obviously, $\true$ represents the value \emph{true}, $\imp$ is implication, and $\all$ stands for universal quantification.

A signature $\sig'$ \deftechterm{extends} a signature $\sig$ if every abstraction of $\sig$ is also an abstraction of $\sig'$, and has the same shape in both $\sig$ and $\sig'$. 

An abstraction logic $\logic'$ \deftechterm{extends} an abstraction logic $\logic$ if the signature of $\logic'$ extends the signature of $\logic$, and if every axiom of $\logic$ is also an axiom of $\logic'$ (modulo $\alpha$-equivalence, see Section~\ref{semantics}).

\section{Several Abstraction Logics}
\deftechterm{Deduction Logic $\logic_D$} has signature $\sig_D$ consisting of the minimally required three abstractions, 
and the following axioms $\axioms_D$:
\begin{description}
\item[\lax{D}{1}] $\ctrue$
\item[\lax{D}{2}] $A \cimplies (B \cimplies A)$
\item[\lax{D}{3}] $(A \cimplies (B \cimplies C)) \cimplies ((A \cimplies B) \cimplies (A \cimplies C))$
\item[\lax{D}{4}] $(\cforall x.\,A[x]) \cimplies A[x]$
\item[\lax{D}{5}] $(\cforall x.\,A \cimplies B[x]) \cimplies (A \cimplies (\cforall x.\, B[x]))$
\end{description}
The relevance of $\logic_D$ is in enabling the \techterm{Deduction Theorem} for abstraction logic (see~\cite{pal}). 

Axioms \lax{D}{2} to \lax{D}{5} might surprise despite their familiar form, because one would expect them to apply only to truth values, not every value in the entire universe. But because abstraction logic has only a single universe $\univ$, every value must also be considered to be a truth value, 
because for \emph{any} term $t$ one can ask the question: 
\begin{center}
``Is $t$ true?''
\end{center}
\noindent\deftechterm{Deduction Logic with Equality $\logic_E$} extends $\logic_D$. It has signature $\sig_E$ which extends $\sig_D$ by adding the equality abstraction $\eq$. It has axioms $\axioms_E$, adding the following axioms to $\axioms_D$:
\begin{description}
\item[\lax{E}{1}] $x \ceq x$
\item[\lax{E}{2}] $x \ceq y \cimplies (A[x] \cimplies A[y])$
\item[\lax{E}{3}] $A \cimplies (A \ceq \ctrue)$
\end{description}

$\logic_E$ is arguably the most important abstraction logic from a theoretical point of view. 
We will see in Section~\ref{completeness} that every AL extending $\logic_E$ is \techterm{complete}.

\absatz\deftechterm{Deduction Logic with Equality and Falsity} $\logic_F$ extends $\logic_E$. Its signature $\sig_F$ extends $\sig_E$ by adding $\false$, denoting \val{false}, logical negation $\cnot$, and inequality $\neq$. Its axioms $\axioms_F$ add the following definitions to $\axioms_E$:
\begin{description}
\item[\lax{F}{1}] $\cfalse \ceq (\cforall x.\, x)$
\item[\lax{F}{2}] $\cnot A \ceq (A \cimplies \cfalse)$
\item[\lax{F}{3}] $(x \neq y) = \cnot(x = y)$ 
\end{description}
From these axioms follows \emph{ex falso quodlibet}, i.e. $\cfalse \imp A$, as well as the \emph{principle of explosion}, i.e. $A \imp (\cnot A \imp B)$.

The possibility of defining $\false$ as $\cforall x.\, x$ has been noted as early as 1951 by Church~\cite{church-sense}, albeit Church let the quantifier range over booleans only. Much later in 2013 Kripke found it remarkable enough to exhibit it in a short note about \techterm{Fregean quantification theory}~\cite{kripke-frege}. Unlike Church, and like us, Kripke lets the quantifier range over the entire universe, which he calls a \techterm{Fregean domain}.

\absatz\deftechterm{Intuitionistic Logic $\logic_I$} extends $\logic_F$. Its signature $\sig_I$ adds abstractions to $\sig_F$ for 
conjunction $\cand$, disjunction $\cor$, equivalence $\cequiv$ and existential quantification $\ex$. 
Its axioms $\axioms_I$ are obtained by adding the following axioms to $\axioms_F$:
\begin{description}
\item[\lax{I}{1}] $(A \cand B) \cimplies A$
\item[\lax{I}{2}] $(A \cand B) \cimplies B$
\item[\lax{I}{3}] $A \cimplies (B \cimplies (A \cand B))$
\item[\lax{I}{4}] $A \cimplies (A \cor B)$
\item[\lax{I}{5}] $B \cimplies (A \cor B)$
\item[\lax{I}{6}] $(A \cor B) \cimplies ((A \cimplies C) \cimplies ((B \cimplies C) \cimplies C))$
\item[\lax{I}{7}] $(A \cequiv B) \ceq ((A \cimplies B) \cand (B \cimplies A))$
\item[\lax{I}{8}] $A[x] \cimplies (\cexists x.\, A[x])$
\item[\lax{I}{9}] $(\cexists x.\, A[x]) \cimplies ((\cforall x.\, A[x] \cimplies B) \cimplies B)$
\end{description}

\absatz\deftechterm{Classical Logic $\logic_K$} extends $\logic_I$ with the \techterm{law of the excluded middle}:
\begin{description}
\item[\lax{K}{}] $A \cor \cnot A$
\end{description}
In classical logic, there are only two logical equivalence classes: $\true$ and $\false$. In particular, the following is a theorem of $\logic_K$:
$$
(A = \true) \cor (A \cequiv \false)
$$

\section{Peano Arithmetic}
As an example, consider how to formulate \techterm{Peano arithmetic}~\cite[section~3.1]{mendelson} as an abstraction logic $\logic_P$. Our starting point is $\logic_I$ or $\logic_K$.
We add the following abstractions to the signature: \emphabs{zero} denotes a value, and both \emphabs{suc} and \emphabs{nat} denote unary operations. 
Here \emphabs{nat} will take on the role of a predicate to determine which elements of the mathematical universe are \techterm{natural numbers}. 
That means we consider $n$ to be a natural number iff \emphabs{nat} applied to $n$ yields \val{true}:
\begin{description}
\item[\lax{P}{1}] $\opnatapp \opzero$
\item[\lax{P}{2}] $\opnatapp n \imp \opnatapp {\opsucapp n}$
\item[\lax{P}{3}] $\opnatapp n \imp (\opsucapp n \neq \opzero)$
\item[\lax{P}{4}] $\opnatapp n \imp (\opnatapp m \imp ({\opsucapp n} = {\opsucapp m} \imp n = m))$
\item[\lax{P}{5}] $P[\opzero] \imp (\all n.\, \opnatapp n \imp P[n] \imp P[\opsucapp n]) \imp \opnatapp n \imp P[n]$
\end{description}
Note that \lax{P}{5} is the \techterm{axiom of induction}. It can be represented in AL effortlessly. 

Addition and multiplication are axiomatised as binary operations \emphabs{add} and \emphabs{mul}:
\begin{description}
\item[\lax{P}{6}] $\opnat n \imp \opaddapp n \opzero = n$
\item[\lax{P}{7}] $\opnat n \cand \opnat m \imp \opaddapp n {\opsucapp m} = \opsucapp {\opaddapp n m}$
\item[\lax{P}{8}] $\opnat n \imp \opmulapp n \opzero = \opzero$
\item[\lax{P}{9}] $\opnat n \cand \opnat m \imp \opmulapp n {\opsucapp m} = \opaddapp {\opmulapp n m} n$
\end{description}
The axioms of $\logic_P$ are conditioned on $n$ and $m$ being natural numbers. This is because the natural numbers share the mathematical universe with other values, for example \val {true}, and this avoids making our axioms stronger than necessary. For example, we might want $\opsucapp \true$ to yield an error value. On the other hand, we might live in a mathematical universe where
$\opzero = \true$, and then $\opsucapp \true$ suddenly makes sense! Another scenario is that we might want to define \emphabs{suc} additionally on booleans $\true$ and $\false$ such that $\opsucapp \true = \false$ and $\opsucapp \false = \true$. Our conditioning ensures that all of these scenarios remain possible.

\section{Undefinedness}
An important aspect of formalisation is how to deal with \techterm{undefinedness}. For our formalisation of Peano arithmetic as an AL, 
we chose to guard axioms such that the operations on natural numbers are \deftechterm{unspecified} outside the domain of natural numbers. Even
the domain of natural numbers itself is unspecified to a large extent, as for example neither $\opnatapp \true$ nor $\cnot \opnatapp \true$ follow from $\logic_{P}$. 

An alternative way of handling undefinedness is to employ \deftechterm{error values}. These are otherwise normal values of the universe meant to signal that an operator does not make sense for a particular combination of arguments, but yields an error instead. To further simplify, one can introduce a single designated error value \val{fail}, denoted by the abstraction $\fail$. To make this work smoothly, a supporting cast of abstractions and axioms is needed, among them: 
\begin{description}
\item[\lax U 1] $(\fail \neq \true) \cand (\fail \neq \false)$
\item[\lax U 2] $(\exu x.\, A[x]) = (\ex x.\, A[x] \cand (\all y.\, A[y] \imp x = y))$
\item[\lax U 3] $(\exu x.\, A[x]) \imp (A[x] \cequiv (\operatorname{the} x.\, A[x]) = x)$
\item[\lax U 4] $\cnot (\exu x.\, A[x]) \imp (\operatorname{the} x.\, A[x]) = \fail$
\end{description}
Here $\exu$ denotes \techterm{unique existence}, and ``$\operatorname{the}$'' denotes the \techterm{definite description operator}. 
Similarly, if one does not shy away from the \techterm{axiom of choice},
it is possible to instead introduce \techterm{Hilbert's choice operator} ``$\operatorname{some}$'' via
\begin{description}
\item[$\text{U}'_3$] $A[x] \imp A[(\operatorname{some} x.\, A[x])]$
\item[$\text{U}'_4$] $\cnot (\ex x.\, A[x]) \imp (\operatorname{some} x.\, A[x]) = \fail$
\end{description}
and then define ``the'' in terms of ``some''.

Farmer proposes a \emph{traditional approach to undefinedness}~\cite{farmer-calculus} as a basis for formalising undefinedness. It allows undefined terms, but formulas cannot be undefined. He mentions that \val{undefined} could be modelled as a third truth value besides \val{true} and \val{false}, but warns that 
\begin{quote}
[...] it is very unusual in mathematical practice to use truth values beyond \val{true} and \val{false}. Moreover, the presence of a third truth value makes the logic more complicated to use and implement. 
\end{quote}
Our approach based on AL and a designated error value resolves these issues. There is a third value $\fail$ with $\fail \neq \true$ and $\fail \neq \false$, but assuming classical logic, 
$\fail$ is logically equivalent to $\false$, because we can prove $\fail \cequiv \false$. Furthermore, there is no need to change the implementation of the logic at all, as introducing additional abstractions and axioms is sufficient. Indeed, our approach is very close to Farmer's traditional approach, but in AL we are not forced to distinguish between formulas and terms. This has the advantage that undefined terms never have to be converted to defined formulas, which can happen in counter-intuitive ways.  

How does one choose between the unspecified and the error value approach to undefinedness? I think these two approaches are not opposing, but rather complementary. 
If the aim is maximal generality and reuse, the unspecified approach has clear advantages. 
On the other hand, if one needs to reason about whether a term $t$ is undefined or not, the error value approach wins, as this can be done by showing $t = \fail$ or $t \neq \fail$.
E.g., it may be useful to know whether $\int_D F(x)\, dx = \fail$ or not. Further special values besides $\fail$ can also make sense, as in $\int_D F(x)\, dx = \infty$. 

In practice we mix both approaches. A formalisation usually consists of one \techterm{big theory}, which is built by consecutive consistency preserving
definitions and proofs of theorems. The development of the big theory is supported by making use of \techterm{libraries} and \techterm{little theories}. The terms big theory and little theory have been coined by Farmer~\cite{farmer-little-theories}. For me, the difference between libraries and a little theory is that a library is used directly, by merging the objects defined in the library into the big theory. On the other hand, little theories are used via \techterm{theory interpretation}, i.e. by mapping objects defined in the little theory to already existing objects in the big theory. Little theories will mainly use the unspecified approach, so that the objects in the little theory can be mapped to many different kinds of existing objects in the big theory. On the other hand, a library will use the error value approach, so that the library consumer can rely on as many guarantees as possible about the imported objects. Of course, imported theories can have aspects of both library and little theory, and any library / little theory can act as a big theory during its development.

\section{Practical Types and Practal}
I have sketched in~\cite{pal} how FOL can be axiomatised in AL. Furthermore I describe in~\cite{aal} how the models of AL
can be axiomatised in HOL, so that existing tools for automating HOL can be repurposed for automating AL. 
It is known that HOL (assuming Henkin semantics) is equivalent to FOL~\cite[corollary 3.6]{shapiro-foundations}. Therefore FOL, HOL and AL all have the same ``strength''.
What makes AL better than FOL and HOL is its simplicity, versatility, and practicality.

These properties of AL are demonstrated and utilised by the mathematical theory of \deftechterm{practical types}, which subsumes HOL and even \techterm{dependent types}. An earlier version of practical types, which actually led to the discovery of abstraction logic, is described in~\cite{practical-types}. This description is now obsolete, and an updated version based on AL is forthcoming. Just as ZF set theory is built on top of predicate logic, practical types are built on top of abstraction logic. 

Practical types themselves have their origin in my vision for \deftechterm{Practal}~\cite{practal}. The idea of Practal is to build an interactive theorem proving system acting as \emph{a true
bicyle for the mathematical mind}. This entailed not starting from a logic, and then building Practal based on the constraints of that logic, but instead letting the demands of Practal shape the logic. This course of action has been a spectacular success, and resulted eventually in abstraction logic. 

There is no prototype of Practal publically available yet. How fast the development of Practal progresses will depend on raising sufficient funding for Practal.

\section{Semantics}\label{semantics}
A \deftechterm{valuation} $\nu$ in the mathematical universe $\univ$ is a function that assigns to each variable $x$ in $X$ occurring with arity $n$ an $n$-ary operation $\nu(x, n)$ of $\univ$. Given a valuation $\nu$ in $\univ$, distinct variables $x_0, \ldots, x_{k-1}$ and values $u_0, \ldots, u_{k-1}$ in $\univ$, we define an updated valuation via 
$$\nu[x_0 \coloneqq u_0, \ldots, x_{k-1} \coloneqq u_{k-1}](y, n) = \begin{cases}u_i & \text{for}\ y = x_i\ \text{and}\ n = 0 \\ \nu(y, n) & \text{otherwise}\end{cases}$$

An \deftechterm{abstraction algebra} is a universe $\univ$ together with a signature $\sig$ and an interpretation $I(a)$ of each abstraction $a$ in $\sig$ as an operator of $\univ$ compatible with the shape of $a$. We say that $a$ \deftechterm{denotes} $I(a)$ (in the given abstraction algebra). 

Abstraction algebras are a generalisation of what Rasiowa calls \techterm{abstract algebras}~\cite{rasiowa}. An \deftechterm{abstract algebra} is an abstraction algebra that only contains abstractions with zero valence. Rasiowa uses abstract algebras as models for propositional calculi, i.e. to provide an \techterm{algebraic semantics} for propositional logics. We take this a step further and use abstraction algebras to provide models for abstraction logics. 

Given such an abstraction algebra, and a valuation $\nu$ in $\univ$, every term $t$ denotes a value $\valueof t$ in the universe $\univ$. The \deftechterm{calculation} of this value is straightforward, and recurses over the structure of $t$:
\begin{itemize}
\item If $t = x$ where $x$ is a variable, then $\valueof t = \nu(x, 0)$.
\item If $t = x[t_0, \ldots, t_{n-1}]$ is a variable application with $n > 0$, then $$\valueof t = f(u_0, \ldots, u_{n-1})$$ where $f = \nu(x, n)$ is the $n$-ary operation assigned to $x$ via the valuation $\nu$, and $u_i$ is the value of $t_i$ with respect to $\nu$. 
\item If $t = (a.)$ then $\valueof t = u$, where $a$ denotes the value $u$.
\item If $t = (a.\, t_0 \ldots t_{n-1})$ for $n > 0$, then $$\valueof t = f(u_0, \ldots, u_{n-1})$$ where $a$ denotes the $n$-ary operation $f$, 
and $u_i$ is the value of $t_i$ with respect to $\nu$. 
\item Assume $t = (a\, x_0\ldots x_{m-1}.\, t_0 \ldots t_{n-1})$ for $m > 0$, and let the shape of $a$ be $(m; p_0, \ldots, p_{n-1})$. 
  Let $F$ be the $n$-ary operator denoted by $a$. We define the arguments $r_i$ to which we will apply $F$ as follows. If $p_i = \emptyset$, then $r_i$ is the value denoted by $t_i$ with respect to $\nu$. If $p_i = \{j_0, \ldots, j_{k-1}\}$ for $\size{p_i} = k > 0$, then the abstraction $a$ \deftechterm{binds} the variables $x_{j_0}, \ldots, x_{j_{k-1}}$ occurring with arity $0$ in $t_i$. Thus $r_i$ is the $k$-ary operation which maps $k$ values $u_0, \ldots, u_{k-1}$ to the value denoted by $t_i$ with respect to the valuation $\nu[x_{j_0} \coloneqq u_0, \ldots, x_{j_{k-1}} \coloneqq u_{k-1}]$. Then $$\valueof t = F(r_0, \ldots, r_{n-1})$$
\end{itemize}

Two terms $s$ and $t$ are called \deftechterm{$\alpha$-equivalent} with respect to a given signature $\sig$ iff for all abstraction algebras $\algebra$ with this signature, and all valuations in (the universe of) $\algebra$, $s$ and $t$ denote the same value. It can easily be checked whether two terms are $\alpha$-equivalent by first computing their \techterm{de Bruijn representations}, and then checking if the two representations are identical. An elaboration and proof of this can be found in~\cite{al} under the moniker ``equivalence of $\alpha$-equivalence 
and \techterm{semantical equivalence}''.

A variable $x$ occurs \deftechterm{free} with arity $n$ in a term $t$ if there is an occurrence of $x$ with arity $n$ in $t$ that is not bound by any surrounding abstraction. Because abstractions bind only variables of arity $0$, \emph{any} occurrence of $x$ in $t$ with arity $n > 0$ is free.

It is clear that the value of $t$ depends only on the assignments in the valuation to those variables $x$ which are free in $t$. Therefore the value of a \deftechterm{closed} term $t$, i.e. a term without any free variables, does not depend on the valuation at all, but only on the abstraction algebra in which the calculation takes place.

\section{Substitution}

An \deftechterm{$n$-ary template} has the form $[x_0 \ldots x_{n-1}.\, t]$,
where the $x_i$ are distinct variables and $t$ is a term. The template binds all free occurrences of $x_i$ in $t$ of arity $0$. A $0$-ary template $[.\, t]$ is usually just written $t$.

A \deftechterm{substitution} $\sigma$ is a function defined on a \techterm{domain} $D$ that maps a variable $x$, given an arity $n$ such that $(x, n)$ belongs to $D$, to an $n$-ary template $\sigma(x, n)$.

The purpose of a substitution $\sigma$ is to be applied to a term $t$, yielding another term $\appsubst{\sigma}{t}$ as the result of the substitution. This is done by determining those free occurrences of $x$ in $t$ of arity $n$ that are in $\sigma$'s domain, and replacing them with $\sigma(x, n)$. This means that expressions of the form $$[x_0 \ldots x_{n-1}.\, t][s_0, \ldots, s_{n-1}]$$ must be resolved to actual terms when replacing variables occurring with non-zero arity. This can in turn be done by applying to $t$ the substitution that maps $x_i$ to $s_i$. 

The usual caveats of substitution in the presence of bound variables apply. For example, assume $\sigma$ maps $x$ to $y$. Then applying $\sigma$ to $(a\, y.\, (b.\, x\, y))$ by performing a direct replacement of $x$ with $y$ would yield $(a\, y.\, (b.\, y\, y))$. This is not what we want,
as can be seen as follows. Note first that $(a\, y.\, (b.\, x\, y))$ is $\alpha$-equivalent to $(a\, z.\, (b.\, x\, z))$, i.e. the meaning of those two terms is the same in all abstraction algebras with respect to all valuations. Therefore it should not matter to which of the two terms we apply the substitution to, the results should be the same, or at least $\alpha$-equivalent. But applying $\sigma$ to the second term (again via direct replacement) yields $(a\, z.\, (b.\, y\, z))$, which is not $\alpha$-equivalent to $(a\, y.\, (b.\, y\, y))$.

These issues can be dealt with by defining substitution via \techterm{de Bruijn terms} instead of terms. The details of this have been worked out in~\cite{al}. For our purposes here we will just note that we can deal with these issues by renaming bound variables appropriately before performing replacement to avoid clashes between free and bound variables. Also note that there is no canonical result of applying a substitution, but that the result is determined only up to $\alpha$-equivalence. 

\subsection*{Substitution as Valuation}

The main property of substitution is that, given a \techterm{background valuation} $\nu$,
we can turn any substitution $\sigma$ into a valuation $\nu_\sigma$. The role of the background valuation is to provide meaning for free variables that are either not in the domain of $\sigma$, or free in some template of $\sigma$.

The valuation $\nu_\sigma$ has the property that for any term $t$
$$
\valueof{t}_\sigma = \valueof{\appsubst{\sigma}{t}}
$$ 
holds. Here $\valueof{\cdot}$ denotes evaluation with respect to $\nu$, and $\valueof{\cdot}_\sigma$ calculates the value with respect to $\nu_\sigma$. 

We define $\nu_\sigma$ as follows for a variable $x$ occurring with arity $n$:
\begin{itemize}
\item If $n = 0$, then $\nu_\sigma(x, n) = \valueof{\appsubst{\sigma}{x}}$.
\item If $n > 0$, then for any values $u_0, \ldots, u_{n-1}$ we define
  $$\nu_\sigma(x, n)(u_0, \ldots, u_{n-1}) = \valueof{t}_{y \coloneqq u}$$ 
  where $\sigma(x, n) = [y_0 \ldots y_{n-1}.\, t]$, and $\valueof{\cdot}_{y \coloneqq u}$ evaluates with respect to the valuation
  $\nu[y_0 \coloneqq u_0, \ldots, y_{n-1} \coloneqq u_{n-1}]$.
\end{itemize}

Further details and a proof of above property using de Bruijn terms can be found in~\cite{al}. 

\section{Models}

What is the meaning of an abstraction logic? 
Terms have been given meaning relative to an abstraction algebra and a valuation earlier in Section~\ref{semantics}. Correspondingly, a \techterm{model} of an abstraction logic $\logic$ with signature $\sig$ and axioms $\axioms$ is \emph{almost} an abstraction algebra $\algebra$ with the same logic signature $\sig$, a universe $\univ$, and an interpretation $I$, such that we have for all valuations $\nu$ in $\univ$ and all axioms $a$ in $\axioms$ $$\valueof{a} = I(\true)$$ 
There are two important modifications to be made before arriving at the actual definition of a model: 
\begin{enumerate}
\item We are not interested in all possible abstraction algebras as models, but only in those that are \techterm{logic algebras}.
\item We are not interested in testing the axioms with all possible valuations, but only with those that belong to a \techterm{space of valuations} depending on the model itself.
\end{enumerate}

To address the first point, we define a \deftechterm{logic algebra} to be an abstraction algebra with logic signature $\sig$ such that $\imp$ and
$\all$ have denotations satisfying the following minimum requirements: 
\begin{itemize}
\item $I(\imp)(I(\true), u) = I(\true)$ implies $u = I(\true)$ for all values $u$ in $\univ$.
\item Let $f$ be any unary operation of $\univ$ such that $f(u) = I(\true)$ for all values $u$ in $\univ$. Then $I(\all)(f) = I(\true)$.
\end{itemize}

To address the second point, we define a \deftechterm{valuation space} $\valuationspace$ of an abstraction algebra $\algebra$ to be a collection of valuations in the universe of $\algebra$ such that:
\begin{itemize}
\item $\valuationspace$ is not empty. 
\item If $\nu$ is a valuation belonging to $\valuationspace$, $u$ is a value in the universe of $\algebra$, 
  and $x$ is a variable in $X$, then $\nu[x \coloneqq u]$ also belongs to $\valuationspace$.
\item If $\nu$ is a valuation belonging to $\valuationspace$, and if $\sigma$ is a substitution with respect to the signature of $\algebra$, 
  then $\nu_\sigma$ also belongs to $\valuationspace$.
\end{itemize}

These requirements for $\valuationspace$ will turn out to be just \emph{strong enough} to show \techterm{soundness} of any abstraction logic, and \emph{weak enough} to prove 
\techterm{completeness} for any abstraction logic extending $\logic_E$.

A \deftechterm{model} of an abstraction logic $\logic$ is then a pair $(\algebra, \valuationspace)$ such that $\algebra$ is a logic algebra with the same signature as $\logic$, and $\valuationspace$ is a valuation space of $\algebra$, such that $\valueof{a} = I(\true)$ for all valuations $\nu$ in $\valuationspace$ and all axioms $a$ in $\axioms$.

A model is called \deftechterm{degenerate} if the universe of $\algebra$ consists of a single value. Note that \emph{every} abstraction logic has a degenerate model. 

If every model of $\logic$ is also a model of the logic with the same signature as $\logic$ and the single axiom $t$, then we say that $t$ is \deftechterm{valid} in $\logic$ and write 
$$\logic \valid t$$

\section{Proofs}

Given a logic $\logic$ with signature $\sig$ and axioms $\axioms$, a \deftechterm{proof} in $\logic$ of a term $t$ is one of the following:
\begin{enumerate}
\item An \deftechterm{axiom invocation} $\pAX(t)$ such that $t$ is an axiom in $\axioms$.
\item An \deftechterm{application of substitution} $\pSUBST(t, \sigma, p_s)$ where 
  \begin{itemize}
  \item $p_s$ is proof of $s$ in $\logic$
  \item $\sigma$ is a substitution 
  \item $t$ and $\appsubst \sigma {s}$ are $\alpha$-equivalent
  \end{itemize}
\item An \deftechterm{application of modus ponens} $\pMP(t, p_h, p_g)$ such that 
  \begin{itemize}
  \item $p_h$ is a proof of $h$ in $\logic$
  \item $p_g$ is a proof of $g$ in $\logic$
  \item $g$ is identical to $h \imp t$
  \end{itemize}
\item \deftechterm{Introduction of the universal quantifier} $\pALL(t, x, p_s)$ where
  \begin{itemize}
  \item $x$ is a variable 
  \item $p_s$ is a proof of $s$ in $\logic$
  \item $t$ is identical to $(\all x.\, s)$
  \end{itemize}
\end{enumerate}

If there exists a proof of $t$ in $\logic$, we say that $t$ is a \deftechterm{theorem} of $\logic$ and write 
$$ \logic \derives t $$

\section{Soundness}

Abstraction logic is \deftechterm{sound}, i.e. every theorem of $\logic$ is also valid in $\logic$:
$$\logic \derives t \quad\text{implies}\quad \logic \valid t$$
We show by induction over $p$ that if $p$ is a proof of $t$ in $\logic$, then $t$ is valid in $\logic$:
\begin{itemize}
\item Assume $p = \pAX(t)$. Then $t$ is an axiom and is therefore valid.
\item Assume $p = \pSUBST(t, \sigma, p_s)$. Then $p_s$ is a proof of $s$, and therefore $s$ is valid. 
  Let $\nu$ be a valuation in $\valuationspace$ for some model $(\algebra, \valuationspace)$ of $\logic$. 
  Then $$\valueof t = \valueof {\appsubst \sigma s} = {\valueof s}_\sigma$$ 
  Because $\nu_\sigma$ is in $\valuationspace$ as well, and $s$ is valid, $${\valueof s}_\sigma =  I(\ctrue)$$
  Together this yields $\valueof t = {\valueof s}_\sigma  = I(\ctrue)$. 
\item Assume $p = \pMP(t, p_h, p_g)$. Then $p_h$ is a proof of $h$, and $p_g$ is a proof of $g$, and therefore 
  both $h$ and $g$ are valid. Let $\nu$ be a valuation in $\valuationspace$ for some model $(\algebra, \valuationspace)$ of $\logic$.
  Then $\valueof h = I(\ctrue)$ and 
  $$
    I(\ctrue) 
    = \valueof g 
    = \valueof {h \imp t} 
    = I(\timplies)(\valueof h, \valueof t) 
    = I(\timplies)(I(\ctrue), \valueof t) 
  $$
  This implies $\valueof t = I(\ctrue)$.
\item Assume $p = \pALL(t, x, p_s)$. Then $p_s$ is a proof of $s$, and therefore $s$ is valid. 
  Let $\nu$ be a valuation in $\valuationspace$ for some model $(\algebra, \valuationspace)$ of $\logic$. 
  Let $\univ$ be the universe of $\algebra$. Then
  $$\valueof t = \valueof {(\tall x.\, s)} = I(\tall)(f)$$
  Here $f$ is a unary operation of $\univ$ where $f(u) = {\valueof s}_{x \coloneqq u}$ for all values $u$ in $\univ$. Because $\nu[x \coloneqq u]$ also belongs to $\valuationspace$ and
  $s$ is valid, we have ${\valueof s}_{x \coloneqq u} = I(\ctrue)$ and therefore $f(u) = I(\ctrue)$ for all $u$. 
  This implies $\valueof t = I(\tall)(f) = I(\ctrue)$.
\end{itemize}

\section{Completeness}\label{completeness}

An abstraction logic $\logic$ is called \deftechterm{complete} if every valid term is also a theorem: 
$$
\logic \valid t \quad \text{implies}\quad \logic \derives t
$$
If $\logic$ extends $\logic_E$, then $\logic$ is complete.

To prove this, we construct the \deftechterm{Rasiowa Model $(\ralg, \valuationspace)$ of $\logic$}, which is a model for $\logic$. The construction relies heavily on the fact that $\logic$ extends $\logic_E$. The universe of the Rasiowa Model consists of equivalence classes of terms, where two terms $s$ and $t$ are called equivalent if $s = t$ is a theorem of $\logic$. This is a variation of a standard approach to prove completeness, originally due to Lindenbaum and Tarski. We write $\rclass t$ for the equivalence class to which a term $t$ belongs. The details of the construction can be found in~\cite{al} (the Rasiowa Model is called Rasiowa Abstraction Algebra there, which is a misnomer). In the construction we make essential use of the fact that $\valuationspace$ does not need to be the class of all valuations, but can be custom tailored as long as we can guarantee that $\valuationspace$ has the properties required of a valuation space.

Most importantly, there exists a valuation in $\valuationspace$ with 
$$\valueof{t} = \rclass{t}$$
for any term $t$, where $\valueof{\cdot}$ evaluates with respect to that valuation. Furthermore, for the interpretation $I$ of $\ralg$ we have $$I(\true) = \rclass \true$$  

From this we can see immediately that $\logic$ is complete. Because assume that $t$ is valid. Then 
$$\rclass t = \valueof t = I(\true) = \rclass{\true}$$
which shows that $t = \true$ is a theorem of $\logic$, and thus $t$ is a theorem as well.

\section{Consistency}

An abstraction logic $\logic$ is called \deftechterm{inconsistent} if all terms are also theorems, and consequently \deftechterm{consistent}
if there is at least one term which is not a theorem.

If $\logic$ is an extension of a deduction logic $\logic_D$, then inconsistency of $\logic$ is equivalent to 
$$\logic \derives \cforall x.\, x$$
This is easy to see. Assume $\cforall x.\,x$ is a theorem in $\logic$. Substituting $[x.\, x]$ for $A$ in Axiom~{\lax D 4} of $\logic_D$ and applying modus ponens, it follows that $x$ is a theorem. Substituting any term $t$ for $x$ shows that $t$ is a theorem.

If an abstraction logic $\logic$ is inconsistent, then all models of $\logic$ are degenerate. Assume $\logic$ is inconsistent. It follows that $x$ is a theorem for any variable $x$ of arity zero. That means that $x$ is valid in any model $(\algebra, \valuationspace)$ of $\logic$. 
In particular, for any valuation $\nu$ in $\valuationspace$ we have $\valueof x = I(\ctrue)$. There is such $\nu$ because $\valuationspace$ is non-empty. For any value $u$ in the universe $\univ$ of $\algebra$ the valuation $\nu[x \coloneqq u]$ is in $\valuationspace$ as well,
and therefore 
$u = \valueof{x}_{x \coloneqq u} = I(\ctrue)$. That means $\univ$ consists only of $I(\ctrue)$, thus any model of $\logic$ is degenerate.

What about the other direction? If every model of an abstraction logic $\logic$ is degenerate, then
$\valueof t = I(\ctrue)$ holds for any term $t$ with respect to any valuation.  
Therefore $$\logic \valid t$$ 
But is $t$ then a theorem of $\logic$? We have seen that this is indeed the case if $\logic$ extends $\logic_E$, because then $\logic$ is complete. 

Therefore, if $\logic$ extends $\logic_E$, then inconsistency of an abstraction logic $\logic$ is equivalent to all of its models being degenerate.

It is thus straightforward to see that $\logic_K$ is consistent. As a model, use a two-element \techterm{boolean algebra} extended in the obvious way with operators denoting $\forall$ and $\exists$. The axioms of $\logic_K$ can then be checked by simply exhausting all possibilities.

\section{Rasiowa's Algebraic Approach}

Abstraction logic is a generalisation of Rasiowa's work on \techterm{algebraic semantics} for nonstandard propositional logics~\cite{rasiowa}. 
Rasiowa chose to generalise her approach to predicate logic by following Mostowski and interpreting quantifiers as least upper bounds and greatest lower bounds. To me, that does not seem to be the proper way to generalise Rasiowa's approach. Instead, I believe abstraction logic is. Just like Rasiowa restricts her models by restricting \techterm{abstract algebras} to 
\techterm{implicative algebras} to ensure that implication satisfies \emph{modus ponens}, AL meets minimal requirements of implication and universal quantification by restricting 
\techterm{abstraction algebras} to \techterm{logic algebras}. Note that \techterm{Heyting algebras} and \techterm{boolean algebras} are special cases of implicative algebras, and provide models for \techterm{intuitionistic propositional logic} and \techterm{classical propositional logic}, respectively.

Why did Rasiowa miss this generalisation, which now seems obvious with hindsight? The reason may simply be that operators were not part of her normal toolbox, while working with higher-order functions is standard in both programming and mechanised theorem proving today. Another possible 
reason is that working with a \techterm{Fregean domain}~\cite{kripke-frege} might have felt alien to her. But on the other hand, her models for propositional logic are topological spaces, and it must have felt somewhat of a waste of space to fill them just with mere truth values. As we can see now, these spaces are meant to represent whole mathematical universes.

Our approach seems to be conceptually simpler and less involved than Rasiowa's approach to predicate logics. An important indicator for this is how simple, elegant and without any need to exclude special cases our completeness proof for abstraction logic is (once the Rasiowa Model is constructed). Completeness holds for all abstraction logics from $\logic_E$ on, including intuitionistic abstraction logic, classical abstraction logic, and all logics in between and beyond. On the other hand, completeness for predicate logics based on algebraic semantics is in general 
complicated territory~\cite{cintula-henkin}. Ono concludes his paper~\cite{ono-completeness} from 1995 on the completeness of predicate logics with respect to algebraic semantics with the words
\begin{quote}
 [...] there seems to be a big difference between instantiations in logic and in algebra. In other words, though usually we interpret quantifiers as infinite joins and meets in algebraic semantics, i.e., we deal with quantifiers like infinite disjunctions and conjunctions, this will be not always appropriate. So, one of the most important questions in the study of algebraic semantics for predicate logics would be how to give an appropriate interpretation of quantifiers, in other words, how to extend algebraic semantics.
\end{quote}
It seems that abstraction logic just might be this extension of algebraic semantics that Ono was looking for.

\section{Further Related Work}

The \emph{Mizar} system~\cite{mizar-nutshell} is a theorem proving system based on predicate logic and set theory. Theories are developed
in the \emph{Mizar language}, which has over 100 keywords, yet has no support for general variable binding. Wiedijk acknowledges that this is an unsatisfying state of affairs, and proposes
an extension of the Mizar language to add general variable binding~\cite{mizar-binders}. It is not clear how practical his proposal is, and it has not been acted upon. 
Mizar has a system of \emph{soft types}~\cite{mizar-soft-types} which are somewhat similar in purpose to \emph{practical types}. 
An important difference is that practical types is a mathematical theory \emph{replacing} set theory, while soft types are mostly syntactic sugar for predicates \emph{on top of} set theory.

\emph{Metamath}~\cite{metamath} is another system mainly used for formalising first-order logic, although its essence is that of a \emph{logical framework}. It does not have a concept of free or bound variables, but has a special instruction for disjoint variables. It is up to the user to wield this instruction properly when implementing a logic. \emph{Metamath Zero}~\cite{metamath-zero} is a modern descendant of Metamath. It also does not distinguish between free and bound variables, but instead allows sort annotations of the form $\phi : \operatorname{wff} x$. This means that in the expression $\phi$ the variable $x$ may occur, but for example the variable $y$ may not. Again, it is up to the user to use this facility responsibly when implementing a logic. 

Most other logical frameworks are based on some weak form of intuitionistic higher-order logic. The one that influenced my work the most is \emph{Isabelle}~\cite{isabelle}. Its meta-logic
\emph{Pure}~\cite{isabelle-foundation} is based on implication, universal quantification, and equality, just as $\logic_E$ is ($\logic_E$ also has $\true$). But like HOL, Pure delegates variable binding to $\lambda$-abstraction. To justify that an object logic built on top of Pure is sound and complete, one has to employ proof-theoretic methods tailored to that specific object logic, 
because Pure is mainly a syntactic mechanism without sufficient semantic foundations. In contrast to this, \emph{every} logic extending $\logic_E$ is guaranteed to be sound and complete.

\section{Conclusion}

Abstraction logic is the \emph{golden middle} between FOL and HOL. On one hand it can be viewed as HOL minus static types. On the other hand it can also be understood as FOL plus operators and, consequently, general variable binding. 

The primary application of AL is to serve as a logical foundation for \emph{interactive theorem proving} and in particular \emph{Practal}~\cite{practal}, but given the simplicity of AL's terms, AL could also be fruitfully applied in 
\emph{computer algebra}. In fact, maybe AL will help to unify those two fields.

I believe that abstraction logic is the best logical foundation for (computer) mathematics. It has the simplest syntax of all general logics: a term is either a \emph{variable application},
or an \emph{abstraction application}. And it has the simplest notion of proofs: a proof is either an \emph{axiom invocation}, a \emph{substitution}, an application of \emph{modus ponens},
or a \emph{universal quantifier introduction}. Still, it is expressive enough to model all three common foundations~\cite[section 1.2]{avigad-foundations} of interactive theorem proving as mathematical theories: FOL, HOL, and dependent types. 

The semantics of abstraction logic is a generalisation of algebraic semantics for propositional logic. As a pleasant consequence there are no vague notions of propositions or predicates involved. Abstraction logic is not complicated philosophy. It is simple mathematics. 

\section{Update}

This paper was submitted to CICM 2022 and rejected (one weak reject, one reject, and one weak accept). I have the feeling the weak accept came from an actual mathematician, so that is encouraging. Apart from that, I don't think the reviewers actually understood what they have in front of them here. But they made important points how this paper could to be improved nevertheless:
\begin{itemize}
  \item ``The paper is too long.'' That is unfortunate, but I am not interested in writing a shorter one at this point. Anyway, it seems the paper in its current form is not getting its point across quickly and clearly enough. 
  \item ``It's full of meta theory. Where are the applications?'' Good point, but I consider the meta theory of AL as a major achievement and breakthrough. So of course I want to present it. Apart from some technical parts of the completeness proof, the meta theory is straightforward and simple. In light of this, I find it even more surprising that AL has not been discovered before. As for applications, I believe Practical Types~\cite{practical-types} and of course Practal~\cite{practal} will be among its major applications, but that is future work.
  \item ``The completeness proof is vague.'' I would not consider the proof vague. It is a proof outline, and for the detailed construction of the Rasiowa Model the reader is referred to~\cite{al}. But I do agree that this is the part of the meta theory that needs to be checked the most carefully. I recently made some observations that make me doubt if Deduction Logic with Equality is actually complete. The construction of the Rasiowa Model is quite technical, and it is very possible that errors crept in. That is why I started to formalise AL and its metatheory
  in Isabelle~\cite{al-in-isabelle}, but this is a project that might take a while.
  \item ``What about Nominal Logic?'' Nominal Logic (NL) is superficially related to AL in that both can be considered extensions of first-order logic, and in that both deal with variable binding. But the aim of NL is to talk \emph{about} syntax, which is very different from AL's purpose. The goal of AL is to be a new foundation for all of logic and mathematics. 
\end{itemize}

\section{Acknowledgements}

Many thanks to Norbert Schirmer for numerous discussions via text chat about Abstraction Logic. Having to justify and explain various points of AL to Norbert really helped me to better understand AL. Also many thanks to Amine Chaieb for reading through \cite{pal} and providing valuable feedback despite him suffering from Covid and headaches during that time! Furthermore, thanks to Mario Carneiro and all other participants of the Zulip chat which spurred my search for Abstraction Logic as a foundation of Practical Types. I would like to thank the anonymous and not so anonymous reviewers of the various journals and conferences that have rejected this paper or an earlier version of it. Their feedback can only make this paper and its future versions better. Finally, the biggest thank you goes to Ania. Without her, Abstraction Logic would most certainly not exist.

\end{document}